%% file: main.tex
\documentclass{article}
\input{dependencies.tex}
\usepackage{fancyvrb}

\usepackage{chngcntr}
\counterwithin{figure}{section}

\title{\bfseries{Trustless Cross-chain Communication for Zendoo Sidechains}}

\author{
    Alberto Garoffolo\\
    \texttt{alberto@horizen.global}\\
    \texttt{Horizen}
    \and
    Dmytro Kaidalov\\
    \texttt{dmytro.kaidalov@iohk.io}\\
    \texttt{Input Output}
    \and
    Roman Oliynykov\\
    \texttt{roman.oliynykov@iohk.io}\\
    \texttt{Input Output}\\
    \texttt{V.N.Karazin Kharkiv National University}
}

\date{September 2022}

\begin{document}

\maketitle
\thispagestyle{empty}

\section*{ \centering }
\begin{center}
    \textbf{Abstract}
\end{center}
In the \textit{Zendoo} white paper \cite{GKO20,Zendoo}, we introduced a novel sidechain construction for Bitcoin-like blockchains, which allows a mainchain to create and communicate with sidechains of different types without knowing their internal structure. In this paper, we take a step further by introducing a comprehensive method for sidechains to communicate amongst each other. We will also discuss the details of a cross-chain token transfer protocol that extends the generic communication mechanism. With the cross-chain token transfer protocol, it can enable a broad range of new applications, such as an \textit{exchange platform}, that allows the ability to trade tokens issued from different sidechains.

\newpage
{
  \hypersetup{linkcolor=black}
  \tableofcontents
}
\newpage
\input{1_Introduction.tex} 
\input{2_Zendoo_overview.tex}
\input{3_Cross_sidechain_comm_protocol.tex}
\input{4_Mitto_token_transfer_protocol.tex}
\input{5_Conclusions.tex}

\bibliographystyle{plain}
\bibliography{references}
\end{document}

%% file: dependencies.tex
\usepackage[utf8]{inputenc}
\usepackage{graphicx}
\graphicspath{ {./images/} }
\usepackage{enumitem}
\usepackage{url}
\usepackage{nameref}

\usepackage[dvipsnames]{xcolor}
\usepackage{hyperref}

\colorlet{mylinkcolor}{violet}
\colorlet{mycitecolor}{YellowOrange}
\colorlet{myurlcolor}{Aquamarine}
\hypersetup{
    breaklinks=true,
	colorlinks=true, %set true if you want colored links
	linktoc=all,     %set to all if you want both sections and subsections linked
    linkcolor=violet,
    filecolor=magenta,      
    urlcolor=cyan,
    citecolor=YellowOrange
}

\usepackage{amsthm}
\theoremstyle{definition}
\newtheorem{definition}{Definition}[section]
\usepackage{geometry}
\newgeometry{vmargin={35mm}, hmargin={35mm,35mm}}

\usepackage{listings}
\usepackage{placeins}
\usepackage{titlesec}
\usepackage{titletoc}
\usepackage[title,titletoc]{appendix}

\usepackage{caption}
\usepackage{array,tabularx}
\newenvironment{conditions}
  {\par\vspace{\abovedisplayskip}\noindent
   \tabularx{\columnwidth}{>{$}l<{$} @{${}-{}$} >{\raggedright\arraybackslash}X}}
  {\endtabularx\par\vspace{\belowdisplayskip}}
\usepackage[skins]{tcolorbox}
\newtcolorbox{protocolframe}[2][]{%
  enhanced,colback=white,colframe=black,coltitle=black,
  sharp corners,boxrule=0.4pt,
  fonttitle=\itshape,
  attach boxed title to top left={yshift=-0.3\baselineskip-0.4pt,xshift=2mm},
  boxed title style={tile,size=minimal,left=0.5mm,right=0.5mm,
    colback=white,before upper=\strut},
  title=#2,#1
}

%% file: 1_Introduction.tex
\section{Introduction}

Since the appearance of Bitcoin \cite{N08}, blockchain technology has gained great attention in both industry and academia. Experts from various fields have started exploring and building new platforms and applications. The distinguishing feature of Bitcoin - the absence of a centralized control - is perceived to be disruptive to the existing financial systems, thereby making them more robust, fair, and transparent. Bitcoin inspired the development of many other similar crypto-platforms with a variety of different features.

The adoption of Bitcoin and other cryptocurrencies grew significantly during the previous decade, which also exposed their limitations such as limited throughput, increased latency, and reduced ability to scale \cite{C16}. In 2014, A. Back et al. proposed a concept of sidechains \cite{BCDF14} that has the potential to overcome these limitations. The basic premise is to create many sub-blockchains that are interoperable with the main blockchain. They operate separately but can transact with the mainchain native asset. In this way, blockchain systems like Bitcoin can extend their functionality implemented in a sidechain (e.g., introduce smart contracts \cite{RSK}).

In 2020, we introduced \textbf{Zendoo} \cite{GKO20,Zendoo} - a universal construction for blockchain systems that allows users to create and communicate with sidechains of different types without knowing their internal structure. Moreover, we proposed a specific sidechain construction, named \textbf{Latus}, that can be built on top of this infrastructure and can realize a decentralized, verifiable blockchain system. It leverages the use of zk-SNARKs \cite{BCTV13,Darlin} to generate succinct proofs of sidechain state progression \cite{Coda,Halo} that are used to validate cross-chain transfers of assets between the mainchain and sidechain.

In this paper, we discuss how sidechains can interoperate with one another. The communication enhances advanced functionalities and enables a plethora of new applications. The development of cross-blockchain communication involves much complexity. Therefore, it is essential to ensure the security and effectiveness of the developed system. The cross-sidechain communication protocol (CSCP) utilizes the features of Zendoo to provide security and avoid the pitfalls of many existing bridging protocols \cite{VB22}. The protocol provides a way for sidechains to transfer messages to each other. 

Finally, we further build on the cross-sidechain communication protocol and extend it for transferring tokens issued on different sidechains. The main  idea is that tokens can be encoded as messages and then transferred in a verifiable method using the CSCP protocol. Thus, transferring of tokens allows users to utilize and trade them on different sidechains, as well as use them in other decentralized financial applications.

The discussion of the cross-sidechain communication protocol for Zendoo is organized in the following manner: 
\begin{itemize}
    \item Section 2 provides a brief overview of Zendoo and provides a detailed description of important components for the communication scheme.
    \item Section 3 provides a detailed description of the generic cross-sidechain communication protocol.
    \item Section 4 introduces the Mitto protocol that extends CSCP and specifies secure token transfers between sidechains.
\end{itemize}

%% file: 2_Zendoo_overview.tex
\section{Zendoo Overview}

The Zendoo construction was introduced in \cite{GKO20}. It allows the deployment of customizable sidechains attached to the Bitcoin-like mainchain. Here we provide a general overview and discuss the portions relevant to the cross-sidechain communication protocol. For a full discussion of Zendoo, see the white paper \cite{Zendoo}.

The main feature of Zendoo is that the mainchain does not know the internal structure of the sidechains. This feature allows it to build a variety of different systems by leveraging the secure and standardized cross-chain communication.
  
\begin{figure}[htbp]
	\centering
	\includegraphics[clip,width=0.9\columnwidth] {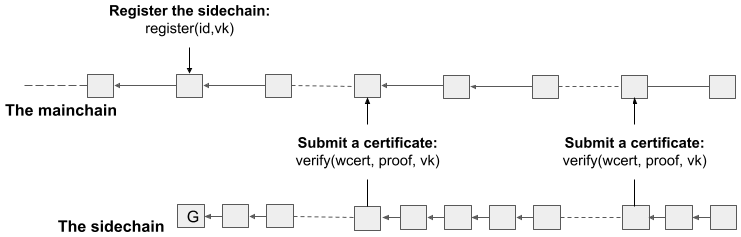}
	\caption{A simplified communication scheme between a sidechain and the mainchain.}
	\label{fig:2_1}
\end{figure}

Zendoo considers the parent-child relationship between the mainchain and sidechains, where sidechain nodes directly observe the mainchain, while mainchain nodes only observe cryptographically authenticated certificates from sidechain maintainers. Certificate authentication and validation are achieved by using SNARKs \cite{BCTV13,Darlin}, which enable constant-sized proofs of arbitrary computations. The main feature of the construction is that sidechains are allowed to define their own SNARKs, thereby establishing their own rules for authentication and validation. All SNARK proofs comply with the same verification interface used by the mainchain which enables universality as the sidechain can use an arbitrary protocol for authenticating its certificates. This basic concept is shown in Fig.~\ref{fig:2_1}.

A sidechain is registered to the mainchain by creating a special transaction, which sets the new sidechain id, its initial state, and more importantly, the SNARK verification key. This is used to validate the certificates. The mainchain expects certificates to be submitted in strict intervals called \textbf{withdrawal epochs}. The length of a withdrawal epoch is also defined in the sidechain creation transaction. 

The basic structure of the withdrawal certificate is defined by the mainchain and this structure must be followed by all sidechains.

\begin{definition}{\textbf{Withdrawal Certificate (WCert).} \label{def:wcert}}
A withdrawal certificate is a standardized posting that allows sidechains (SC) to communicate with the mainchain (MC). Its main functions are:
\begin{itemize}
    \item delivering backward transfers to the MC
    \item serving as a heartbeat message and enabling the mainchain to identify the sidechain status.
\end{itemize}
WCert is represented by a tuple of the form:
\[WCert\ \stackrel{\mathrm{def}}{=}\ (ledgerId,\ epochId,\ quality,\ BTList,\ proofdata,\ proof),\]
where,
\begin{conditions}
    ledgerId & an identifier of the sidechain for which WCert is created, \\
    epochId &  a number of a withdrawal epoch, \\
    quality & an integer value that indicates the quality of this withdrawal certificate, \\
    BTList & a list of backward transfers included in this withdrawal certificate, \\
    proofdata & input data to a SNARK verifier, \\
    proof & a SNARK proof.
\end{conditions}
\end{definition}

As previously mentioned, the certificate is verified upon submission to the mainchain. The SNARK verification interface is the same for all sidechains:
\[ true/false \leftarrow Verify(vk_{WCert},\ public\_input,\ proof),\]
\vspace*{-6mm}
\[ public\_input\ \stackrel{\mathrm{def}}{=}\ (wcert\_sysdata,\ MH(proofdata)),\]
where,
\begin{conditions}
    vk_{WCert} & a SNARK verification key registered upon the sidechain creation; \\
    wcert\_sysdata &  a part of the public input, which is unified for all sidechains and enforced by the mainchain (explained further); \\
    proofdata & a part of the input data defined by the sidechain and passed along the withdrawal certificate; it is a list of variables of predefined types whose semantics are not known to the mainchain; \\
    MH(proofdata) & a root hash of a Merkle tree where leaves are variables from proofdata; it is essential for the SNARK to keep a list of public inputs short. Thus we combine them in a tree and pass the root hash only\footnote{A full payload of \textit{proofdata} is provided during the proof generation as a witness.}; \\
    proof & a SNARK proof itself submitted as a part of the certificate.
\end{conditions}

The \textit{wcert\_sysdata} parameter plays an important role from a security standpoint. The idea is to allow the mainchain to verify the proof against some public input parameters defined by the protocol. For instance, the \textit{BTList} and \textit{quality} parameters that are part of the certificate must be verified before being used by the mainchain.

The \textit{wcert\_sysdata} is represented by the tuple of the following form:
 
\[ wcert\_sysdata\ \stackrel{\mathrm{def}}{=}\ (quality,\ MH(BTList)),\ B(_{last}^i)),\]
where,
\begin{conditions}
    quality & the quality parameter from the withdrawal certificate, \\
    MH(BTList) &  a root hash of a Merkle tree where leaves are backward transfers from the BTList provided within the certificate, \\
    H(B_{last}^i) & a block hash of the last mainchain block in the withdrawal epoch $i$.
\end{conditions}

The flexibility of the scheme comes from the ability of a sidechain to define \textit{proofdata}. This can be an arbitrary set of arguments completely defined by the sidechain. For instance, it can contain a commitment of the entire state of the sidechain at the moment of creating the certificate. Then, the SNARK proof proves the state transition from the state committed in the previous certificate.

\subsection{Ceased sidechain} \label{sec:ceased_sc}

As previously stated, the sidechain is obliged to submit a withdrawal certificate in strict intervals called \textbf{withdrawal epochs}. If a withdrawal certificate has not been submitted during this time, the sidechain is considered \textbf{ceased} and no more withdrawal certificates for this sidechain will be accepted by the mainchain. However, the funds can still be withdrawn from the ceased sidechain by means of a \textbf{ceased sidechain withdrawal}. Since the certificates are not allowed for the ceased sidechains, it becomes the backup method to retrieve funds.

\begin{definition}{\textbf{Ceased Sidechain Withdrawal (CSW).} \label{def:CSW}}
The CSW is an operation that allows the movement of coins from the ceased sidechain \textit{B} to the original mainchain \textit{A}. It is represented by a tuple of the following form:
\[CSW\ \stackrel{\mathrm{def}}{=}\ (ledgerId,\ receiver,\ amount,\ nullifier,\ proofdata,\ proof),\]
where,
\begin{conditions}
    ledgerId & an identifier of the sidechain for which CSW is created, \\
    receiver &  an address of the receiver in the mainchain, \\
    amount & the number of coins to be transferred, \\
    nullifier & a unique identifier of claimed coins, \\
    proofdata & input data to a SNARK verifier, \\
    proof & a SNARK proof.
\end{conditions}
\end{definition}

The CSW is submitted to the mainchain and has the same validation principle as the withdrawal certificate. It is validated by a SNARK proof defined by the sidechain. The verification key $vk_{csw}$ for the proof is set upon sidechain registration. The syntax of the \textit{proofdata} and proof are the same as for the withdrawal certificate. The basic interface of the SNARK verifier is the following:

\[ true/false \leftarrow Verify(vk_{csw},\ public\_input,\ proof),\]
\vspace*{-6mm}
\[ public\_input\ \stackrel{\mathrm{def}}{=}\ (csw\_sysdata,\ MH(proofdata)),\]
where,
\begin{conditions}
    vk_{csw} & a SNARK verification key for the CSW registered upon the sidechain creation; \\
    csw\_sysdata &  a part of the public input, which is unified for all sidechains and enforced by the mainchain (explained further); \\
    proofdata & a part of the input data defined by the sidechain and passed along the CSW; it is a list of variables of predefined types whose semantics are not known to the mainchain; \\
    MH(proofdata) & a root hash of a Merkle tree where leaves are variables from proofdata; \\
    proof & a SNARK proof itself submitted as a part of the CSW.
\end{conditions}

\textit{csw\_sysdata} is defined as:
\[ csw\_sysdata\ \stackrel{\mathrm{def}}{=}\ (H(B_w),\ nullifier,\ receiver,\ amount),\]

where $H(B_w)$ is a block hash of the mainchain block where the latest withdrawal certificate for this sidechain has been submitted.
\\~\\
It is completely up to the sidechain to define the semantics of the \textit{proofdata} (i.e., what type of data it contains) and verification logic. For instance, one might require proof that a user, who creates CSW, owns a UTXO holding the \textit{amount} of coins in the state committed by the last certificate before the sidechain was ceased.

See the Zendoo white paper \cite{Zendoo} for more information on ceased sidechain withdrawals. 

\subsection{Sidechain transactions commitment} \label{sec:sc_tx_comm}

\begin{figure}[htbp]
	\centering
	\includegraphics[trim={0.5cm 8.6cm 8cm 1.6cm}, clip,width=0.9\columnwidth] {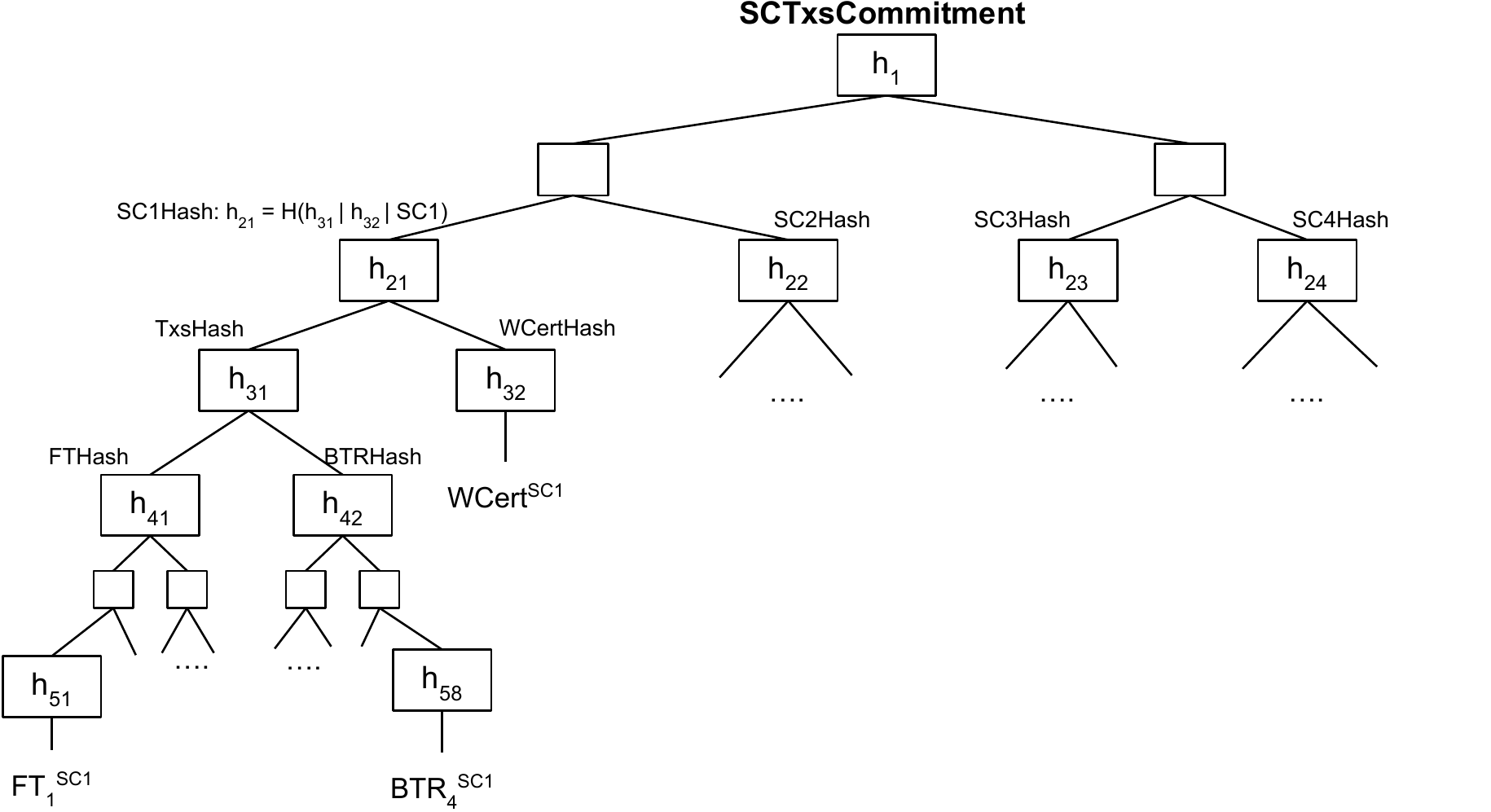}
	\caption{Sidechain transactions commitment tree.}
	\label{fig:2_2}
\end{figure}

An important concept for the cross-chain communication in Zendoo is the \textbf{Sidechain Transactions Commitment (STC)}. The STC is a special value that is inserted in every mainchain block and comprises all sidechain-related actions in the MC block for all registered sidechains. The STC value is a root hash of a Merkle tree that contains all transactions related to any sidechain  (see Fig.~\ref{fig:2_2}). 

The root hash $h_1$  commits to all sidechain related transactions included in the MC block. All $SC\textit{\textbf{X}}Hash$ ($h_{2X}$), where $X$ is a sidechain identifier, are ordered by its id and commit to all transactions related to the sidechain $X$. $WCertHash$ commits to the WCert for the sidechain $X$ (if present) and only one WCert is allowed for each sidechain. $TxsHash$ commits to different types of transactions for a particular sidechain.
\\~\\
The STC allows access to all sidechain-related information. Recall that the interface of SNARK verification for both WCert and CSW requires that the input to the proof contains MC block hash (Def.~\ref{def:wcert}-\ref{def:CSW}). This means that a SNARK proof can access the entire history of all sidechain transactions by proving that a certain STC with a certain SC transaction is included in some MC block. The connection of such MC block to the one provided in the proof can be proved recursively as the MC blocks are cryptographically chained to each other.

%% file: 3_Cross_sidechain_comm_protocol.tex
\section{Cross-Sidechain Communication Protocol}

The Zendoo construction defines only the communication protocol between the mainchain and a sidechain. It focuses mostly on forward and backward transfers of the native MC asset. A desirable feature is the ability of disparate sidechains to communicate with one another. It allows for the creation of powerful new functionality that leverages the benefits of multiple sidechain networks, simultaneously. However, such cross-sidechain communication can not be a part of the Zendoo protocol itself because the core concept postulates that the internal structure of the sidechains is unknown to Zendoo.

\begin{figure}[htbp]
	\centering
	\includegraphics[width=0.6\columnwidth] {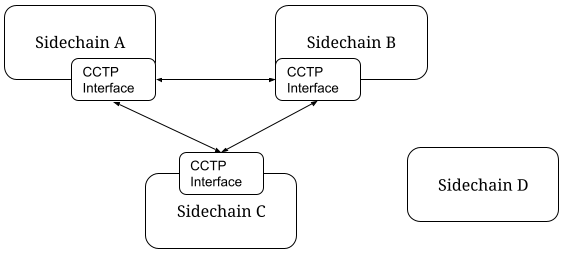}
	\caption{Sidechains A,B,C implement a special CCTP Interface that allows them to communicate with each other. Sidechain D does not implement the CCTP interface.}
	\label{fig:3_1}
\end{figure}

Therefore, to enable cross-sidechain communication we introduce a separate \textbf{Cross-Sidechain Communication Protocol (CSCP)} that extends Zendoo and defines how the sidechains communicate with one another. The basic idea is that sidechains that want to interact with each other must implement a special generalized interface (see Fig.~\ref{fig:3_1}).

The CSCP defines an abstract mechanism of verifiable transferring of messages from one sidechain to another. It is implemented by a sidechain using customization capabilities of Zendoo. The mainchain and Zendoo itself are not affected because Zendoo already has all the necessary capabilities to create a custom logic inside the sidechains. 

\begin{figure}[htbp]
	\centering
	\includegraphics[width=0.5\columnwidth] {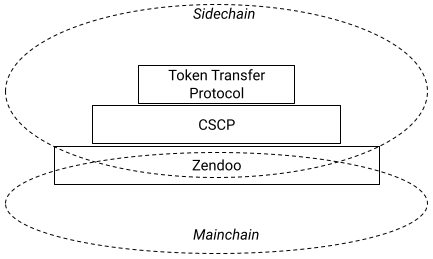}
	\caption{A layered structure of communication protocols. Zendoo specifies the basic protocol for communication between the mainchain and a sidechain. CSCP relies on Zendoo and specifies the abstract protocol for communication amongst the sidechains. Next, the Token Transfer protocol is built on top of CSCP and specifies a concrete protocol for exchanging tokens among sidechains.}
	\label{fig:3_2}
\end{figure}

Then, the sidechains define particular types of messages that are exchanged along with the processing rules for them. For instance, they might implement custom token transfers based on CSCP interface (see Fig.~\ref{fig:3_2}).

\subsection{Message passing}

In a nutshell, the CSCP protocol works the following way: given two sidechains $SC_1$ and $SC_2$, if a user in $SC_1$ wants to send a message $M_i$ to the sidechain $SC_2$, he creates a transaction $tx(M_i)$ containing $M_i$ and submits it to $SC_1$. At the end of a withdrawal epoch all cross-sidechain messages are collected in a Merkle tree, the root hash of the tree becomes a part of the sidechain state that is included in the withdrawal certificate. When the certificate is confirmed in the mainchain, the state cannot be reverted. Therefore, once the state of $SC_1$ with $tx(M_i)$ has been committed and confirmed in the mainchain, the message can be securely confirmed in $SC_2$. A user in $SC_2$ can receive a message by directly observing $SC_1$, extracting the message, and submitting a transaction $tx(M_i,proof)$ to $SC_2$ together with the proof that the message $M_i$ has been processed in $SC_1$ and is present in the state confirmed in the withdrawal certificate of $SC_1$.

\begin{samepage}
There are  three main phases of the CSCP protocol: 
\begin{enumerate}[leftmargin=5em, itemsep=0em]
    \item \textbf{Sending} the message in the sender sidechain.
    \item \textbf{Committing} of the message in the mainchain.
    \item \textbf{Receiving} the message in the receiver sidechain.
\end{enumerate}
\end{samepage}

\begin{figure}[htbp]
	\centering
	\includegraphics[width=0.9\columnwidth] {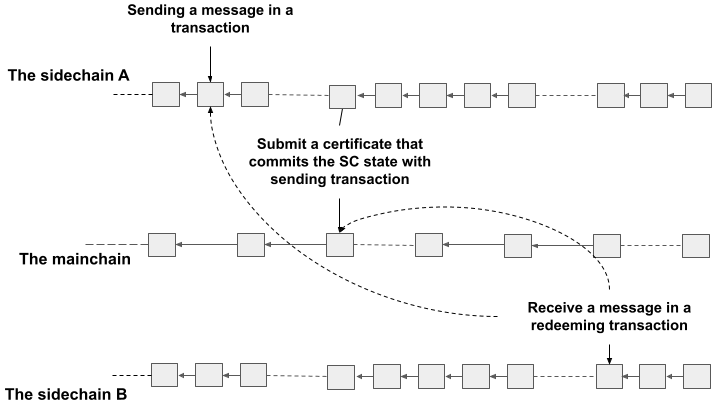}
	\caption{Main phases of the CSCP protocol. The receiving transaction relies on the sending transaction and commitment of the Sidechain A in the mainchain.}
	\label{fig:3_3}
\end{figure}

The main concept is depicted in Fig.~\ref{fig:3_3} while the following subsections provide more details about different phases of the protocol.

\subsubsection{Sending a message}

The CSCP message is defined as follows:
\begin{protocolframe}{}
\begin{lstlisting}
type CSCP_Message {
     sendingScId:    Int,
     receivingScId:  Int,
     msgType:        Int,
     senderId:       ByteArray,
     receiverId:     ByteArray,
     payloadHash:    ByteArray
}
\end{lstlisting}
\end{protocolframe}
\noindent
where,
\begin{conditions}
    sendingScId & the id of the sending sidechain as it was registered in the mainchain;\\
    receivingScId & the id of the receiving sidechain as it was registered in the mainchain;\\
    msgType & the type of the message (e.g., token);\\
    senderId & the unique id of the sender. The semantics of this id is not known to the CSCP protocol and is defined by the implementation of the sending sidechain (e.g., public key);\\
    receiverId & the unique id of the receiver. The semantics of this id is defined by the receiving sidechain (e.g., public key);\\
    payloadHash & the hash of the message itself.
\end{conditions}

The structure of the CSCP message is generalized, where the payload of the message is represented as a hash, which means that the message can be anything depending on a particular instantiation of the CSCP protocol. For instance, the message  can be some entity like a token, transaction output, or a text string.

We assume that on the sending sidechain, a user can create a special transaction which contains \textit{CSCP\_Message}. For instance, in the UTXO-based sidechain the message can be represented as a special output.

\subsubsection{Committing a message}

\begin{figure}[htbp]
	\centering
	\includegraphics[width=0.9\columnwidth] {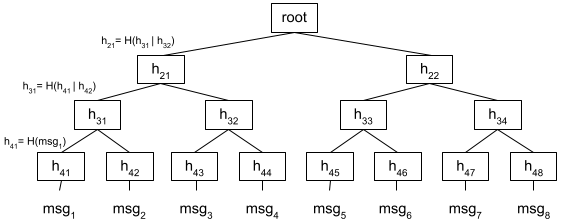}
	\caption{An example of the Merkle tree with CSCP messages. It contains eight messages, but the tree expands depending on the actual number of messages in a particular epoch.}
	\label{fig:3_4}
\end{figure}

At the end of the withdrawal epoch in the sending sidechain, all CSCP messages are combined into a Merkle tree (Fig.~\ref{fig:3_4}).

The messages ($msg_i$) are inserted in the tree in the order they are submitted to the sidechain. The tree contains messages from different users and to different sidechains. Next, the root hash of the tree is inserted in the withdrawal certificate at the end of the epoch as a custom field in the \textit{proofdata} section. The certificate proof enforces the validity of the CSCP messages tree.

\subsubsection{Receiving a message} \label{sec:receive_msg}

To receive a message, a user on the receiving sidechain has to create a special redeem transaction. To do this, the user directly observes the sending sidechain for messages directed to him, extracts the message, and constructs a special transaction that includes both the message and the proof of its validity:

\begin{protocolframe}{}
\begin{lstlisting}
CSCP_Redeem_Tx: {
    msg: 		CSCP_Message,
    payload: 	ByteArray,
    proof: 	    ByteArray	
}
\end{lstlisting}
\end{protocolframe}
\noindent
where \textbf{\textit{msg}} is the original CSCP message from the sending sidechain, the \textbf{\textit{payload}} is a data represented by \textit{\textbf{payloadHash}}, and \textbf{\textit{proof}} is some abstract proof that verifies the following statements about message validity:
\begin{itemize}
    \item The message \textbf{\textit{msg}} has been accepted and confirmed in the sending sidechain.
    \item The \textit{\textbf{payload}} corresponds to \textit{\textbf{payloadHash}} included in \textit{\textbf{msg}}.
    \item The hash of a Merkle tree of CSCP messages containing \textit{\textbf{msg}} is committed in the withdrawal certificate of the sending sidechain and the certificate has been confirmed in the mainchain.
    \item The message \textit{\textbf{msg}} has not been redeemed in the receiving sidechain.
    \item The user who redeems the message, has rights to do so.
\end{itemize}

The CSCP protocol does not define the exact instantiation of the proof. Depending on the sidechain implementation there might be different ways to implement it. For instance, if the sidechain is based on Latus construction \cite{Zendoo}, then its blocks contain references to the MC blocks, which as discussed in section~\ref{sec:sc_tx_comm}, contain \textit{Sidechain Transaction Commitment Tree} (STC). Given the access to MC blocks, the sidechain can access STC values, which commit to all SC-related transactions in all sidechains. Therefore, it can also prove cryptographically that the certificate of the sending sidechain with the message was included in the mainchain.

The side effect of processing a received message is defined completely by the SC logic.

\subsection{Message passing from a ceased sidechain}

Under certain circumstances there might be a need to send a message from a ceased sidechain. For instance, if the sidechain has tokens that a user wants to recover in another sidechain, then a message can be passed. The CSCP protocol defines a mechanism based on the \textit{Ceased Sidechain Withdrawal} (recall CSW from Def.~\ref{def:CSW}) that allows “to withdraw message” from the ceased sidechain.

In this case, the protocol works in the following way: let there be a ceased sidechain $SC_1$ and a user $U_1$ who owns some entity $E_1$ (e.g., a token) that is withdrawable by CSW according to the rules of the sidechain $SC_1$. The CSW SNARK proof of $SC_1$ is implemented in the way that allows verification of the validity of the \textit{CSCP\_Message} provided in \textit{proofdata} as a custom field. A user creates the CSW on the mainchain to withdraw the entity $E_1$ and inserts the message $M_1$ in \textit{proofdata}. Once the CSW is confirmed in the mainchain, the message can be safely received in $SC_2$. A user in $SC_2$ can receive a message by directly observing the mainchain, extracting the message, and submitting a transaction $tx(M_1,proof)$ to $SC_2$ together with the proof that the message $M_1$ has appeared in the CSW for $SC_1$.

The primary objective of the CSW mechanism, as introduced in the Zendoo white paper \cite{Zendoo}, is to recover MC native assets that have been sent to the sidechain. However, it can also be used for recovering abstract messages as well. In this case, the \textit{amount} field of the CSW has to be zero, which means that there is a recovery of something other than the native asset.

%% file: 4_Mitto_token_transfer_protocol.tex
\section{\textit{Mitto} Token Transfer Protocol}

One of the key concepts, when considering blockchain interoperability, is the ability to transfer assets among different chains. There are many considerations, but the main use cases are:
\begin{itemize}
    \item allow the ability to trade assets from different chains,
    \item enable users to borrow assets on one sidechain by providing tokens issued on another sidechain as collateral,
    \item enable transfer of liquidity among different sidechains.
\end{itemize}

In this section we introduce \textit{Mitto} - an extension to the CSCP protocol that enables transferring of tokens issued on different sidechains. The basic idea is that tokens can be encoded as messages and transferred in a verifiable way using the CSCP protocol. The sent tokens are burned on the sending side, while on the receiving side they are redeemed when receiving the message.

One of the main design principles of the Mitto Token Transfer Protocol is that tokens are tracked by the issuing sidechain when they are sent. Therefore, it is impossible for a malicious sidechain to send back forged tokens that have not been previously transferred there. This also implies that the tokens can only be sent from or to the original sidechain, but not among “foreign” sidechains. 

Sidechains that enable token transfer protocol must implement certain interfaces and adhere to rules of processing tokens that are specified in this section. The Mitto protocol does not rely on a specific sidechain construction and can be integrated into different sidechains given that they follow the general rules of processing tokens. The Mitto protocol can be viewed as something similar to ERC-20 or ERC-721 standards in Ethereum. For unification purposes, the protocol supports both fungible and non-fungible tokens.

\subsection{Tokens data structures and state}

The core of the \textit{Mitto} Token Transfer Protocol is the concept of \textit{TokenInstance}. It is an entity that is used to represent tokens inside the sidechain. The protocol does not specify how exactly token instances are managed by the sidechain, it is only assumed that every sidechain that enables \textit{Mitto} holds tokens as \textit{TokenInstanc}es and follows certain rules when transferring them. The \textit{TokenInstance} has the following structure:

\begin{definition}{\textbf{Token Instance.} \label{def:token_instance}}
\textit{TokenInstance} is a data structure that represents a single non-fungible token or some amount of fungible tokens from the same set identified by a name.  It is defined as follows:
\begin{protocolframe}{}
\begin{lstlisting}
type TokenInstance {
    tokenName:      String,
    fungibility:    Boolean
    Union {
        tokenId:    Int,	// for non-fungible tokens
        amount:     Int, 	// for fungible tokens
    }
    issuerSidechainID:  ByteArray,
    ownerPubKey:        ByteArray,
    dataHash:           ByteArray,
}
\end{lstlisting}
\end{protocolframe}
\noindent
where,
\begin{conditions}
    tokenName & a unique name that identifies a set of fungible or non-fungible tokens; \\
    
    fungibility &  indicates whether the tokens specified by \textit{tokenName} are fungible or not. Depending on this flag, the \textit{TokenInstance} structure contains either \textit{tokenId} (for non-fungible tokens) or \textit{amount} field (for fungible tokens). The \textit{fungibility} flag is the same for all token instances with the same \textit{tokenName}; \\
    
    tokenId & if the token is non-fungible, then the \textit{tokenId} defines a unique identifier of a particular token in the set defined by the \textit{tokenName}. In this case \textit{TokenInstance} represents a single token with a unique \textit{tokenId}; \\
    
    amount & if the token is fungible, then the \textit{amount} represents an amount of fungible tokens held by the token instance; \\
    
    issuerSidechainID & the unique identifier of the sidechain where the tokens with tokenName were originally issued; \\
    
    ownerPubKey & the public key of the owner of a particular token instance; \\
    
    dataHash & the hash of other token-related data that is not relevant to the \textit{Mitto} protocol. Recall that the \textit{TokenInstance} is an abstract data structure which defines the basic information needed to manage tokens. A specific token type is defined by the sidechain logic and contains additional data representing the token.
\end{conditions}
\end{definition}

The tokens can be fungible and non-fungible. A fungible token can be represented by several token instances that have the same \textit{tokenName}, but different \textit{amounts}. A set of non-fungible tokens can be represented by several token instances with the same \textit{tokenName}, but different \textit{tokenIDs}, where each token instance represents a unique non-fungible token from the set. The sidechain can issue several different sets of fungible or non-fungible tokens. 

\begin{figure}[htbp]
	\centering
	\includegraphics[width=0.9\columnwidth] {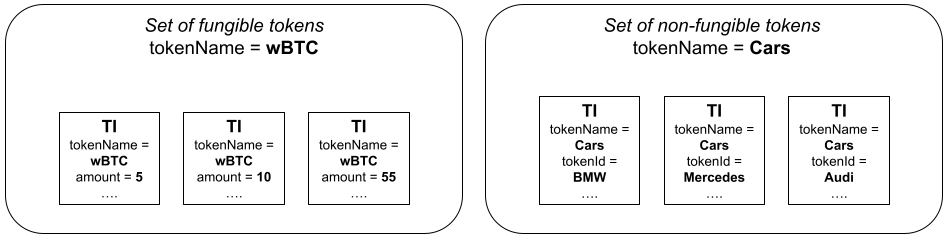}
	\caption{Example of sets of fungible (wBTC) and non-fungible (Cars) tokens. \textbf{TI} is a \textit{TokenInstance}. A sidechain may issue different sets of token instances, it also can contain token instances issued on other sidechains.}
	\label{fig:4_1}
\end{figure}

The \textit{TokenInstance} structure basically resembles the concept of an output in a UTXO based blockchain. But in \textit{Mitto} it is represented abstractly so that the protocol can be implemented also in a non-UTXO based blockchain. 

Another data structure that is required to manage tokens in sidechains is \textit{SentRecord}, which allows to keep track of tokens that have been sent outside the sidechain.

\begin{definition}{\textbf{Sent Record.} \label{def:send_record}}
\textit{SentRecord} is a data structure that keeps track of the tokens that were originally issued on the sidechain but then have been sent to some other sidechain. It is represented as follows:
\begin{protocolframe}{}
\begin{lstlisting}
type SentRecord {
        receiverSidechainID: Int,
        tokenName:           String,
        fungibility:         Boolean
        Union {
            tokenID:    Int, 	// for non-fungible tokens
            amount:     Int,  	// for fungible tokens
        }
}
\end{lstlisting}
\end{protocolframe}
\noindent
where,
\begin{conditions}
    receiverSidechainID & unique identifier of the sidechain where the tokens of type \textit{tokenName} were sent; \\
    
    tokenName &  a unique name that identifies a set of fungible or non-fungible tokens; \\
    
    fungibility & defines whether the tokens specified by \textit{tokenName} are fungible or not. Depending on this flag, the \textit{SentRecord} structure contains either \textit{tokenId} (for non-fungible tokens) or \textit{amount} field (for fungible tokens); \\
    
    tokenId & if the token is non-fungible, \textit{tokenId} is an identifier of a particular token that has been sent outside the sidechain of origin. In this case \textit{SendRecord} represents a single token with a unique \textit{tokenId}; \\
    
    amount & if the token is fungible, \textit{amount} represents an amount of fungible tokens of type \textit{tokenName} that have been sent to sidechain \textit{receiverSidechainID}.
\end{conditions}
\end{definition}

The \textit{TokenInstance} and \textit{SentRecord} are basic structures used for token management in the \textit{Mitto} token transfer protocol. It is assumed that a sidechain keeps track of both \textit{TokenInstances}, that represent the tokens that a sidechain currently possesses, and \textit{SentRecords}, that represent the tokens that have been sent outside the sidechain where they were issued. Let formalize the sidechain state as follows:
\[SC_i^{state}=\{S_i^{tks},S_i^{sent},...\},\]
where,
\begin{itemize}
    \item $S_i^{tks}$ is a set of token instances that are currently owned by the sidechain $SC_i$;
    \item $S_i^{sent}$ is a set of \textit{SentRecords} to keep track of tokens that were issued on sidechain $SC_i$ but were sent outside it. Each \textit{SentRecord} from the set keeps track of the amount of tokens of a particular type that were sent to a particular sidechain.
\end{itemize}

Note that $S_i^{tks}$ may contain token instances representing tokens issued on other sidechains (i.e., $issuerSidechainID \neq SC_i$).
	
The following sections formalize the rules for transferring tokens using the model described above.

\subsection{Sending tokens}

The tokens are sent in the form of CSCP messages which wrap token instances. To send tokens from $SC_a$ to $SC_b$ a user creates a special sending transaction $tx^{send}_{SC_a}(TI_t \rightarrow M_t)$ on $SC_a$ which burns a token instance $TI_t$ and produces an CSCP message $M_t$ with the following data:

\begin{protocolframe}{}
\begin{lstlisting}[mathescape]
$M_t$: CSCP_Message = {
    sendingScId 	= $SC_a$,
   	receivingScId 	= $SC_b$,
   	msgType 		= $TokenTransfer$,
   	senderId 		= $TI_t.ownerPubKey$,
   	receiverId 	    = receiverPubKey,
   	payloadHash 	= $Hash(TI_t)$
}
\end{lstlisting}
\end{protocolframe}
\noindent
where $receiverPubKey$ is the receiver’s public key in $SC_b$ specified by a user. Note that the validity of the transaction (i.e., that the user is the owner of $TI_t$) is verified upon its processing by the sidechain.

We formalize the process of sending tokens by defining the validation rules for the $tx^{send}_{SC_a}(T_I \rightarrow M_t)$ and the side effects of it to the state of $SC_a$ in case the transaction is valid and accepted by the sidechain $SC_a$.

\begin{protocolframe}{\textbf{Validation rules}}
\begin{enumerate}
    \item Verify $TI_t \in S_a^{tks}$ (i.e., the token instance exists in the state of $SC_a$).
    \item If $TI_t.issuerSidechainId \neq SC_a$, then verify that \\ $M_t.receivingScId == TI_t.issuerSidechainId$ (i.e., verify that foreign tokens are sent only to their original blockchain and not elsewhere).
    \item Verify semantic validity of $M_t$:
    \begin{enumerate}
        \item $M_t.sendingScId == SC_a$;
        \item $M_t.receivingScId \neq SC_a$;
        \item $M_t.msgType == TokenTransfer$;
        \item $M_t.senderId == TI_t.ownerPubKey$;
        \item $M_t.payloadHash == Hash(TI_t)$.
    \end{enumerate}
    \item Verify that the transaction is authorized by the owner of $TI_t$ (i.e., there is a signature on $tx^{send}_{SC_a}(TI_t \rightarrow M_t)$ under the public key $TI_t.ownerPubKey$).
\end{enumerate}
\end{protocolframe}

\begin{protocolframe}{\textbf{Side effects on the state of $SC_a$}}
\begin{enumerate}
    \item $TI_t$ is removed from $S_a^{tks}$;
    \item If $TI_t.issuerSidechainId == SC_a$ (i.e., the token issued on this sidechain is being sent), then:
    \begin{enumerate}
        \item If $TI_t.fungibility == false$, then create a sent record \\
        \begin{lstlisting}[mathescape]
            $SR_t$: SentRecord = { 
	            receiverSidechainID = $M_t.receivingScId$,
	            tokenName = $TI_t.tokenName$,
	            fungibility = $false$,
	            tokenId = $TI_t.tokenId$
	        }
        \end{lstlisting}
    and add $SR_t$ to $S_a^{sent}$ (the sidechain keeps track of all tokens that were issued by it and sent elsewhere);
    \item If $TI_t.fungibility == true$, then check if $S_a^{sent}$ already contains some sent record $SR_t^{'}$ such that $SR_t^{'}.receivingSidechainID == M_t.receivingScId$ and $SR_t^{'}.tokeName == TI_t.tokenName$. If yes, just update \\ $SR_t^{'}.amount = SR_t^{'}.amount + TI_t.amount$, otherwise create a sent record \\
    \begin{lstlisting}[mathescape]
            $SR_t$: SentRecord = { 
	            receiverSidechainID = $M_t.receivingScId$,
	            tokenName = $TI_t.tokenName$,
	            fungibility = $false$,
	            amount = $TI_t.amount$
	        }
    \end{lstlisting}
    and add $SR_t$ to $S_a^{sent}$ (the sidechain keeps track of the total amount of tokens that were sent to every other sidechain).
    \end{enumerate}
\end{enumerate}
\end{protocolframe}

At the end of the epoch all CSCP messages are collected in the Merkle tree and the hash of the tree is included in the withdrawal certificate as a custom field, so that they can be received in another sidechain.

\subsection{Receiving tokens} \label{sec:mitto_receive_tokens}

To receive a token from $SC_a$ in $SC_b$ a user creates a special redeem transaction $tx_{SC_b}^{redeem}(M_t \rightarrow TI_t, proof)$ on $SC_b$ that includes the message $M_t$, that has been previously sent from $SC_a$, the corresponding token instance $TI_t$, and a proof of the message validity. The redeem transaction is an extension of the $CSCP\_Redeem\_Tx$ (see section \ref{sec:receive_msg}). It has the following structure:

\begin{protocolframe}{}
\begin{lstlisting}[mathescape]
type Mitto_Redeem_Tx {
    $M_t$:      CSCP_Message,
    $proof$:    ByteArray,
    $TI_t$:     TokenInstance
}
\end{lstlisting}
\end{protocolframe}
\noindent
where $M_t$ is the message that has been previously confirmed in $SC_a$ and that wraps the token instance $TI_t$. The proof proves the validity of the message.

As in the case with tokens sending, we formalize the process of receiving by defining the validation rules for the redeem transaction and the side effects to the state of $SC_b$ in case the transaction is valid and accepted by the sidechain $SC_b$.

\begin{protocolframe}{\textbf{Validation rules}}
\begin{enumerate}
    \item Verify $TI_t.issuerSidechainId == M_t.sendingScId$ or $TI_t.issuerSidechainId == SC_b$ (it is allowed to receive tokens only from the sidechains that issued them or the tokens that were created by the receiving sidechain).
    \item If $TI_t.issuerSidechainId == SC_b$, then verify that:
        \begin{enumerate}
        \item If $TI_t.fungibility == true$, then there exists some sent record $SR_t^{'}  \in S_b^{sent}$ such that:
            \begin{enumerate}
                \item $SR_t^{'}.receivingSidechainID == M_t.sendingScId$, and
                \item $SR_t^{'}.tokenName == TI_t.tokenName$, and
                \item $SR_t^{'}.fungibility == true$, and
                \item $SR_t^{'}.amount \geq TI_t.amount$ (i.e., if we receiving back the native token then it should be present in $S_b^{sent}$ and the $amount$ should not be more than has been originally sent to that sidechain).
            \end{enumerate}
        If such $SR_t^{'}$ does not exist, reject the transaction.
        \item If $TI_t.fungibility == false$, then there should exist some sent record $SR_t^{'} \in S_b^{sent}$ such that:
            \begin{enumerate}
                \item $SR_t^{'}.receivingSidechainID == M_t.sendingScId$, and
                \item $SR_t^{'}.tokenName == TI_t.tokenName$, and
                \item $SR_t^{'}.fungibility == false$, and
                \item $SR_t^{'}.tokenId == TI_t.tokenId$ (i.e., if we receiving back an NFT token from the $tokenName$ set, then it should be present in $S_b^{sent}$).
            \end{enumerate}
        If such $SR_t^{'}$ does not exist, reject the transaction.
        \end{enumerate}
    \item If $TI_t.fungibility == false$ then verify that $TI_t \notin S_b^{tks}$ (if it is an NFT, it can not already be present in the receiving sidechain).
    \item Verify semantic validity of $M_t$:
        \begin{enumerate}
            \item $M_t.sendingScId  \neq SC_b$;
            \item $M_t.receivingScId == SC_b$;
            \item $M_t.msgType == TokenTransfer$;
            \item $M_t.senderId == TI_t.ownerPubKey$;
            \item $M_t.payloadHash == Hash(TI_t)$.
        \end{enumerate}
    \item Verify that the message $M_t$ is authorized by the sender (previous owner of $TI_t$) (i.e., there is a signature on $M_t$ under the public key $M_t.senderId$).
    \item Verify that the redeem transaction is authorized by the receiver (the new owner of $TI_t$) on $SC_b$ (i.e., there is a signature on $tx_{SC_b}^{redeem}(TI_t \rightarrow M_t, proof)$ under the public key $M_t.receiverId$).
    \item Verify the proof, which should check the validity of the message according to CSCP rules (section \ref{sec:receive_msg}).
\end{enumerate}
\end{protocolframe}

Note that the \textit{Mitto} token transfer protocol does not specify precise rules for the message validity proof. Similarly to the CSCP protocol, we view \textit{Mitto} as an abstract protocol that can be integrated into sidechains with different architecture, different state management, etc. That is why at this level of abstraction it is impossible to precisely define the verification logic. For instance, for the Latus-based sidechains the proof can be a SNARK proof that verifies existence of the message in the confirmed certificate of $SC_a$. 

\begin{protocolframe}{\textbf{Side effects on the state of $SC_b$}}
\begin{enumerate}
    \item Let $TI_t^{new}$ be a copy of $TI_t$ except that $TI_t^{new}.ownerPubKey = M_t.receiverId$.
    \item $TI_t^{new}$ is added to $S_b^{tks}$.
    \item If $TI_t^{new}.issuerSidechainID == SC_b$ (i.e., we received back the token that was issued in this sidechain), then:
        \begin{enumerate}
            \item If $TI_t^{new}.fungibility == false$, then find such $SR_t \in S_b^{sent}$ such that $SR_t.receivingSidechainID == M_t.sendingScId$, and $SR_t.tokenName == TI_t.tokenName$, and $SR_t.tokenId == TI_t.tokenId$ and remove it from $S_b^{sent}$;
            \item If $TI_t^{new}.fungibility == true$, then find such $SR_t \in S_b^{sent}$ such that $SR_t.receivingSidechainID == M_t.sendingScId$ and $SR_t.tokenName == TI_t.tokenName$. Then update $SR_t.amount = SR_t.amount - TI_t.amount$. If, after update, $SR_t.amount == 0$, then remove $SR_t$ from $S_b^{sent}$.
        \end{enumerate}
\end{enumerate}
\end{protocolframe}

\subsection{Transferring tokens from a ceased sidechain}

By transferring tokens from a ceased sidechain we understand the case when some sidechain $SB_a$ has been ceased while holding some token instances $TI_i \in S_a^{tks}$ or storing sent records $SR_i \in S_a^{sent}$ for tokens issued by $SC_a$ and sent elsewhere, and a user wants to transfer such tokens. In a nutshell, this can be done by utilizing the Ceased Sidechain Withdrawal mechanism. A user should create a CSW containing a message with the transferred token instance. The SNARK proof of the CSW should be designed in a way that allows to prove the existence of such a TI in the state of the sidechain at the moment of ceasing. Then, the token can be redeemed in the receiving sidechain by creating a special CSW redeem transaction:

\begin{protocolframe}{}
\begin{lstlisting}[mathescape]
type Mitto_CSW_Redeem_Tx {
    $M_t$:      CSCP_Message,
    $proof$:    ByteArray,
    $TI_t$:     TokenInstance,
    $CSW_t$:    CSW
}
\end{lstlisting}
\end{protocolframe}

The validation rules and side effects are almost similar to the one described in section \ref{sec:mitto_receive_tokens} for regular redemption. The only exception is the proof, which in this case should verify that the message has been committed in the $CSW_t$ rather than in a withdrawal certificate. The side effects remain completely the same as in the case of \textit{Mitto\_Redeem\_Tx}.

While receiving is relatively simple, the token sending from a ceased sidechain is more complex. We consider two basic scenarios of withdrawing tokens from a ceased sidechain:
\begin{enumerate}
    \item \textbf{Withdrawing native tokens} (i.e., tokens that were issued on the ceased sidechain).
    \item \textbf{Withdrawing foreign tokens} (i.e., tokens that were not issued on the ceased sidechain, but were held there at the moment of ceasing).
\end{enumerate}

\subsubsection{Withdrawing native tokens}

Let assume that the sidechain $SC_a$ has ceased. There are two cases to consider:
\begin{enumerate}
    \item At the moment of ceasing the sidechain $SC_a$ was the owner of the token $TI_i$ to be withdrawn. I.e., there is a token instance $TI_i \in S_a^{tks}$, where $S_a^{tks}$ is a set of owned tokens at the moment of ceasing. In this case, it is needed to create a CSW that proves the existence of $TI_i$ in $S_a^{tks}$ and the right to withdraw it. The CSW is submitted to the mainchain. Once it is confirmed in the mainchain, \textit{Mitto\_CSW\_Redeem\_Tx} can be used to redeem the token in another sidechain.
    
    \item At the moment of ceasing the sidechain $SC_a$ did not hold the token to be withdrawn. I.e., there is a sent record $SR_t \in S_a^{sent}$, that represents a token(s) that were originally issued in $SB_a$ but currently is owned by some sidechain $SC_c$. We assume that $SC_c$ is active. Such tokens can not be directly transferred from $SC_c$ to another active sidechain $SC_b$, because according to the \textit{Mitto} protocol foreign tokens can be sent only to the sidechain where they have been originally issued. Therefore, to transfer such tokens, $SC_c$ first has to send them back to the ceased sidechain $SC_a$ using the standard sending flow. Then, they can be withdrawn using the CSW mechanism. The CSW should be designed to prove the following:
    \begin{enumerate}
        \item there exist a valid message $M_t$ that were sent by $SC_c$ to $SC_a$, the message has been committed to the confirmed withdrawal certificate of $SC_c$;
        \item the message $M_t$ transfers the token instance $TI_t$, which represent a token(s) that were originally issued by $SC_a$;
        \item there is a sent record $SR_t \in S_a^{sent}$ that confirms that the token(s) have been sent to $SC_c$ previously and that it has not been withdrawn yet.
    \end{enumerate}
    Once such CSW is confirmed in the mainchain, \textit{Mitto\_CSW\_Redeem\_Tx} can be used to redeem the token in the sidechain $SC_b$.
\end{enumerate}

\subsubsection{Withdrawing foreign tokens}

Let assume that the sidechain $SC_a$ has been ceased and it owns a token instance $TI_t \in S_A^{tks}$, such that $TI_t.issuerSidechainId \neq SC_a$. In this case it is needed to create a CSW that proves the existence of $TI_t$ in $S_a^{tks}$ and the right to withdraw it. The CSW will include a corresponding message $M_t$ that wraps $TI_t$ as a custom field. Then, $TI_t$ can be redeemed with \textit{Mitto\_CSW\_Redeem\_Tx} in sidechain $TI_t.issuerSidechainId$ (and only there).

\subsection{Design rationale}

One of the main design principles of the \textit{Mitto} token transfer protocol is that tokens (in the form of token instances) can only be sent from the sidechain where they have been originally issued to some other sidechain and back. But it is disallowed to send tokens from the sidechain where they have not been issued to another sidechain, which is also not their original place of issuance. 

More formally: if there is a message $M_t$ that sends a token instance $TI_t$ from sidechain $SC_a$ to $SC_b$, then
\begin{itemize}
    \item either $SC_a == TI_t.issuerSidechainId$,
    \item or $SC_b == TI_t.issuerSidechainId$.
\end{itemize}

The reason for such a design decision is security. Recall that a sidechain that issues some set of tokens keeps track (through \textit{SentRecord} mechanism) of tokens that have been sent outside. And when these tokens return back they are checked through \textit{SentRecords}. In this case, it is impossible for some corrupted sidechain to send back more tokens than was initially sent there by the issuer. This feature is similar to the \textit{Withdrawal Safeguard} feature introduced in Zendoo (see section 4.1.2.2 in \cite{Zendoo}).

Let's consider this situation in more detail and assume that it is allowed to send foreign tokens from the sidechain where they have not been issued to another third-party sidechain. Let there be several fungible token instances $TI_i$ of type $TI_t.tokenName = SuperCoin$, such that $TI_i.issuerSidechainId = SC_a$. Let assume they are held by $SC_b$ at the moment (i.e., $TI_t \in S_b^{tks}$) and $SC_b$ sends some of them to $SC_c$. But given that this operation is outside $SC_a$, it can not track this transfer and update sent records accordingly (decrease balance of $SC_b$ and increase balance of $SC_c$). Possible solutions to this problem can be:
\begin{enumerate}
    \item Eliminate recording the receiving sidechain in the sent record and just save the amount of tokens of a particular type that have been sent outside. But this creates a vulnerability: if the sidechain $SC_c$ becomes corrupted, such that it can forge the sending transaction to pretend to transfer more tokens that it really has, then the original $SC_a$ has no way to detect this and will accept a malicious transfer that may drain $SR.amount$ to zero and prevent $SC_b$ from transferring back the valid $SC_a$ tokens it holds.
    \item Eliminate the concept of sent records at all, but then a malicious sidechain would be able to forge any amount of tokens and send them back to the original sidechain without any control.
    \item Notify the original sidechain $SC_a$ every time when tokens are transferred somewhere. I.e., if $SC_b$ sends $TI_t$ to $SC_c$, then an additional transaction is created on $SC_a$ that updates $SR$ accordingly. Theoretically, it solves the problem, but it will require to design and implement complex set of proofs and it would be basically equivalent to sending tokens through $SC_a$. We do not consider this option due to its complexity.
\end{enumerate}

There are some other solutions as well, but they are even more complex.

\subsubsection{Token wrapping}

An interesting concept that could also serve as a workaround for the mentioned above problem is token wrapping. For instance, if a sidechain $SC_b$ holds some tokens $TI_i$ issued in $SC_a$, where $TI_i.tokenName=ACoin$ and $TI_i.issuerSidechainId=SC_a$, it can issue a new set of tokens $TI_j$, such that $TI_j.tokenName=wACoin$ and $TI_j.issuerSidechainId=SC_b$. The idea is that $wACoin$ tokens can serve as a wrapper for $ACoin$ tokens. Then, $SC_b$ can send $TI_j$ to other sidechains.

Wrapping may be particularly interesting for the native mainchain asset. Given that every sidechain may contain a certain amount of native MC assets, representing them in the form of wrapped tokens can simplify interaction among different sidechains.

%% file: 5_Conclusions.tex
\section{Conclusions}

We presented a generic protocol for cross-sidechain communication for Zendoo sidechains. It allows different chains to exchange messages with one another in a verifiable way. Moreover, we presented an extension of the protocol that allows to transfer tokens issued on different sidechains.

We view our work as a base for building interoperable Zendoo sidechains. The protocol does not impose any rules on the sidechains structure and can be used by different systems with different security models. 